# Effects of Magnetic Focusing on Power and Stability of VE Amplifiers

**Vadim J. Jabotinski, Alexander N. Vlasov, Simon J. Cooke**

U.S. Naval Research Laboratory
Washington, DC 20375, USA

**Abstract:** Theory and modeling of new discovered effects of the oscillating *e*-beam boundary formed in a magnetic focusing channel on the stability, gain, and power of VE amplifiers are presented. The RF structure and *e*-beam circuit parameters are computed for different beam envelope radii and oscillation amplitudes and then the self-excitation thresholds are obtained from the Determinant equations. A new solution of the beam envelope equation not limited by the small-amplitude oscillations is derived and used to determine the magnetic focusing conditions providing various beam envelope radii and oscillation amplitudes. It is shown for the example $K_a$-band serpentine structure that the optimum radius as well as axial position of the oscillating beam envelope allow achieving significantly greater gain and power of RF amplifiers.

**Keywords:** circuit parameters; electron beam; envelope equation, magnetic focusing; power; rf structure; stability

## Introduction

Attainable power of vacuum electronic, VE, amplifiers is often limited by unwanted self-excitation, SE, that occurs even under zero-drive conditions due to multiple feedbacks and RF coupling between the gaps in such multigap TWT, if the *e*-beam current exceeds the SE-threshold [1]. It was shown [2] that solenoidal magnetic, *B*-, field affects the SE current thresholds and thus the gain and power.

This paper describes for the first time the effects of the oscillating beam boundary, formed in a magnetic focusing channel, on the SE thresholds and attainable gain and power. In addition to the effects of the beam envelope radius and its oscillations the material presents a new effect of phase resonance amplification with respect to the *e*-beam axial position. Also, a new analytical solution of the beam envelope equation accurate for any-amplitude oscillations is shown. For finding the SE current thresholds and frequencies, one first computes and analyzes the structure $Z_1$ and beam $Y_2$ circuit parameter matrices with the Determinant equations [1]. The structure $Z_1$-matrix is determined using the Z-matrix joining formulae [3]. The *e*-beam $Y_2$-matrix is obtained following [2] for variable *B*-focusing conditions, which include the solenoidal *B*-fields in the focusing channel and on-cathode as well as the injection initial beam radius and slop. The *B*-focusing conditions are determined employing the new beam envelope solution (1) given below, to realize a broad range of the beam radii and oscillation amplitudes subjected in this study. In the next step, for the structure defined by the abovementioned $Z_1$ and $Y_2$ matrices, the RF amplifier output signal and gain are calculated by formulation [2], which is confirmed by the EM PIC simulations.

## Effect of the e-beam radius and oscillating boundary on the self-excitation current threshold

Calculations conducted for the example $K_a$-band serpentine structure, Fig. 1, show the SE current thresholds getting smaller for the larger average radius beam in both cases of the small $\Delta r/\bar{r} = 0.01$ (red line) and large $\Delta r/\bar{r} = 0.88$ (blue line) beam radius variation. The small variation model has the structure immersed in uniform axial *B*-field. The large variation model has different on-cathode *B*-field.

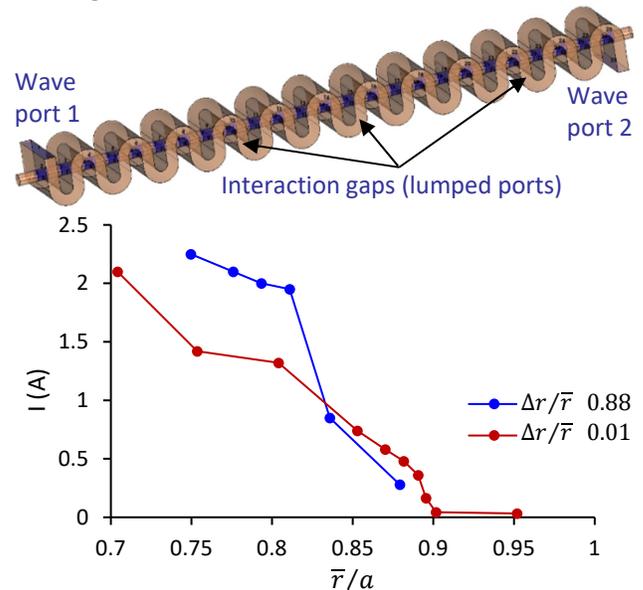

Fig. 1. An example multigap structure (top) and SE current thresholds computed for the 20kV *e*-beam of variable average radius $\bar{r}$ in $a$=0.4mm radius beam tunnel of a 25-gap-long $K_a$-band serpentine structure.

## Attainable gain and power

The trend of the SE thresholds, seen in Fig. 1, to decrease with the larger beam radius suggests that the optimum of radius $\bar{r}$ providing maximum gain and power should be smaller than the beam tunnel radius but sufficiently large for efficient interaction with the EM wave. Such behavior with the maximum gain is demonstrated by the plots in Fig. 2.

## Effect of the phase of the oscillating e-beam boundary – Phase resonance amplification

Axial position of the *e*-beam with oscillating boundary with respect to the gaps of the slow wave structure can affect the SE thresholds and interaction efficiency with the EM wave. In this regard the beam envelope with the maxima positioned inside the gaps would result in a higher power gain as compared to the case of the beam envelope minima positioned inside the gaps. This new phenomenon of phase resonance amplification is demonstrated by the results shown in Fig. 3. The applied *B*-focusing conditions provide the beam envelope oscillation wavelength equal to the RF structure gap-to-gap distance and $\Delta r/\bar{r} = 0.88$.

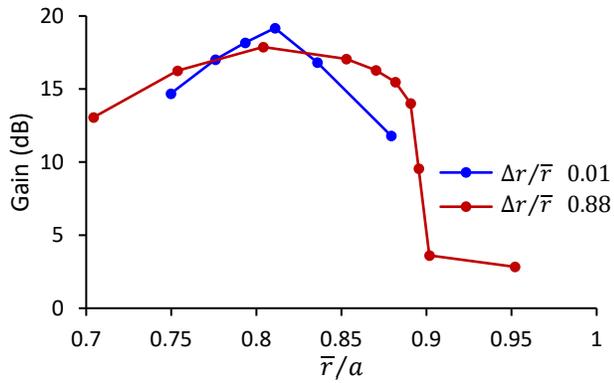

Fig. 2. Attainable power gain computed for the $K_a$-band serpentine 25-gap structure with the 20-kV beam voltage and the current set by the SE thresholds found for various beam radii, oscillation amplitudes, and related *B*-fields.

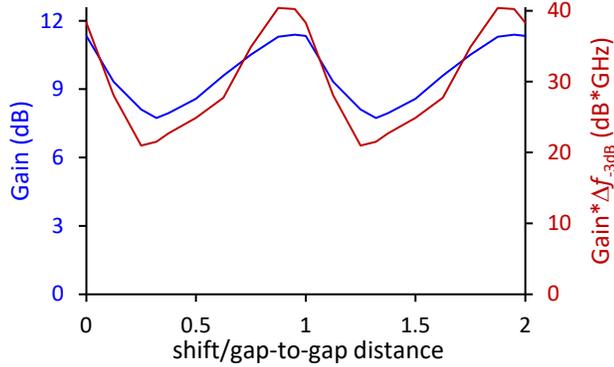

Fig. 3. Power gain (blue line) and gain-bandwidth (red line) computed for various axial positions (shifts) of the 20kV 1.8A *e*-beam with oscillating beam envelope in the *B*-focusing channel in the $K_a$-band serpentine 25-gap structure. The beam current is set below the 1.95A SE threshold found for the selected *B*-fields.

### Solving the particle beam envelope equation

To model a broad range of the beam radii and oscillation amplitudes in the performed studies, the required beam injection and *B*-focusing conditions are determined from the beam envelope equation [4]. The equation assumes the paraxial approximation conditions. Its known solution uses the small parameter method, which imposes additional limits on the radial oscillations. Such small amplitude solution describes the beam envelope by a sinusoid. It however does not hold for larger oscillation amplitudes. The obtained new analytical solution of the beam envelope equation is valid for any level of the beam envelope oscillation amplitude. The solution method is to transform the beam envelope equation to the fist-order integral and apply the Taylor series centered at the roots of the integrand functions split in two intervals. Unlike the small oscillation solution, the solution found accurately describes the squared beam envelope radius as a piecewise sinusoid.

To introduce the new solution and definitions note that the reduced order envelope equation is given by $\frac{z}{r_0} = \int_1^x \frac{dx/2}{\sqrt{y(x)}}$ where $z$ is the particle, $e^-$, beam propagation distance, $r_0$ and $r_0'$ are the initial value and derivative of the beam envelope radius $r(z)$, $x = (r/r_0)^2$, and

$$y(x) = -k_0^2 r_0^2 x^2 + \left(r_0'^2 + k_0^2 r_0^2 + \frac{\theta^2}{r_0^2}\right)x - \frac{\theta^2}{r_0^2} + Kx \ln x$$

Here $k_0 = \frac{c}{2E_0}\frac{B}{\gamma\beta}$ is the wavenumber for the beam envelope with the Lorentz factor $\gamma$ and velocity ratio $\beta$ to the speed of light $c$ in vacuum, in the magnetic focusing field $B$, $K = \frac{1}{2\pi\varepsilon_0 c E_0}\frac{I}{(\gamma\beta)^3}$ is the generalized perveance of the beam with current $I$, $\theta^2 = \varepsilon^2 + \left(\frac{c}{2\pi E_0}\frac{\psi_0}{\gamma\beta}\right)^2$ is a term representing repulsive effects due to the beam emittance $\varepsilon$ and initial angular momentum, on-cathode magnetic flux $\psi_0$, $E_0$ is the electron rest energy, and $\varepsilon_0$ is the vacuum permittivity.

The derived beam envelope solution is given by

$$x = x_0 + \begin{cases} (x_2 - x_o)\sin(2\sqrt{k_2}z + \varphi) &, x \geq x_o \\ |x_1 - x_o|\sin(2\sqrt{k_1}z + \varphi) &, x \leq x_o \end{cases} \quad (1)$$

Here with the auxiliary parameters $\varphi_1$, $\varphi_2$, and $\kappa$

$$\varphi_1 = \sin^{-1}\left(\frac{1-x_0}{x_0-x_1}\right) \quad \varphi_2 = \sin^{-1}\left(\frac{1-x_0}{x_2-x_0}\right) \quad \kappa = \sqrt{k_1/k_2}$$

for the special case $r_o \geq \bar{r} \quad r_0' = 0$

$$\varphi = \begin{cases} \varphi_2 &, x \geq x_o, x_{x_o}' \leq 0 \\ \pi - \varphi_2\kappa &, x \leq x_o \\ -\varphi_2 - \pi/\kappa &, x \geq x_o, x_{x_o}' \geq 0 \end{cases}$$

where $x_0$ is a splitting point at the maximum of $y(x)$, $x_1$ and $x_2$ are the roots of $y(x)$, $k_1$ and $k_2$ are the coefficients of the quadratic terms of the Taylor series of $y(x)$ taken with a positive sign, and $\bar{r}$ is the average beam radius. In addition to the above special case, the solutions are defined for all six possible special cases each depending on the beam injection initial conditions. The obtained solution (1) is in full agreement with the electrostatic and EM PIC simulations and the MATLAB numerical ODE solver.

This work was supported by the Office of Naval Research. Distribution Statement A: Approved for public release, distribution is unlimited.